\documentclass{IEEEcsmag}

\usepackage[colorlinks,urlcolor=blue,linkcolor=blue,citecolor=blue]{hyperref}
\expandafter\def\expandafter\UrlBreaks\expandafter{\UrlBreaks\do\/\do\*\do\-\do\~\do\'\do\"\do\-}
\usepackage{upmath,color}
\usepackage[super,sort]{natbib}

\usepackage{mdframed}
\usepackage[normalem]{ulem}

\jvol{XX}
\jnum{XX}
\paper{8}
\jmonth{Month}
\jname{Publication Name}
\jtitle{Publication Title}
\pubyear{2021}

\begin{document}

\sptitle{Preprint - Submission under review}

\title{Legal Aspects for Software Developers Interested in Generative AI Applications}

\author{Steffen Herbold}
\affil{University of Passau, Passau, Germany}

\author{Brian Valerius}
\affil{University of Passau, Passau, Germany}

\author{Anamaria Mojica-Hanke}
\affil{University of Passau, Passau, Germany}

\author{Isabella Lex}
\affil{University of Passau, Passau, Germany}

\author{Joel Mittel}
\affil{University of Passau, Passau, Germany}

\markboth{THEME}{THEME}

\begin{abstract}
Recent successes in Generative Artificial Intelligence (GenAI) have led to new technologies capable of generating high-quality code, natural language, and images. The next step is to integrate GenAI technology into products, a task typically conducted by software developers. Such product development always comes with a certain risk of liability. Within this article, we want to shed light on the current state of two such risks: data protection and copyright. Both aspects are crucial for GenAI. This technology deals with data for both model training and generated output. We summarize key aspects regarding our current knowledge that every software developer involved in product development using GenAI should be aware of to avoid critical mistakes that may expose them to liability claims. 
\end{abstract}

\maketitle

\chapteri{G}enerative AI (GenAI) technologies changed from research prototypes to products available on the market that receive much attention, e.g., for the generation of code (e.g.,GitHub co-pilot), natural language responses (e.g., ChatGPT), and multi-modal models that combine text and visual information like Dall-E and Midjourney. Due to the opportunity to support or even replace time-consuming manual labor, there is an enormous pressure to bring products to the market, meaning that this domain is currently transitioning from a research-focused prototype generation to an industry-focused tool market.

At the core of this transition are the software developers who create such tools. They need guidance not only on how to use and the capabilities of GenAI tools but also on the legal aspects of their use: when tools are published and/or marketed, there is always a risk of liability. Within this article, we outline key legal arguments regarding two aspects that affect any GenAI technology: data protection and copyright. 

To make the abstract legal concept accessible, we present the legal analysis in a use-case-driven manner. First, we define scenarios that may occur when working with GenAI. Second, we will describe the legal implications for developers that stem from these scenarios. For these considerations, we differentiate between the GenAI models as the learned weights of a neural network and the GenAI applications as something offered as a service, e.g., an API or end-user application.  Further, our considerations are based on laws and regulations from the European Union and the USA. As a result, we derive five lessons for developers using GenAI. 

\begin{figure*}
\centerline{\includegraphics[width=\textwidth]{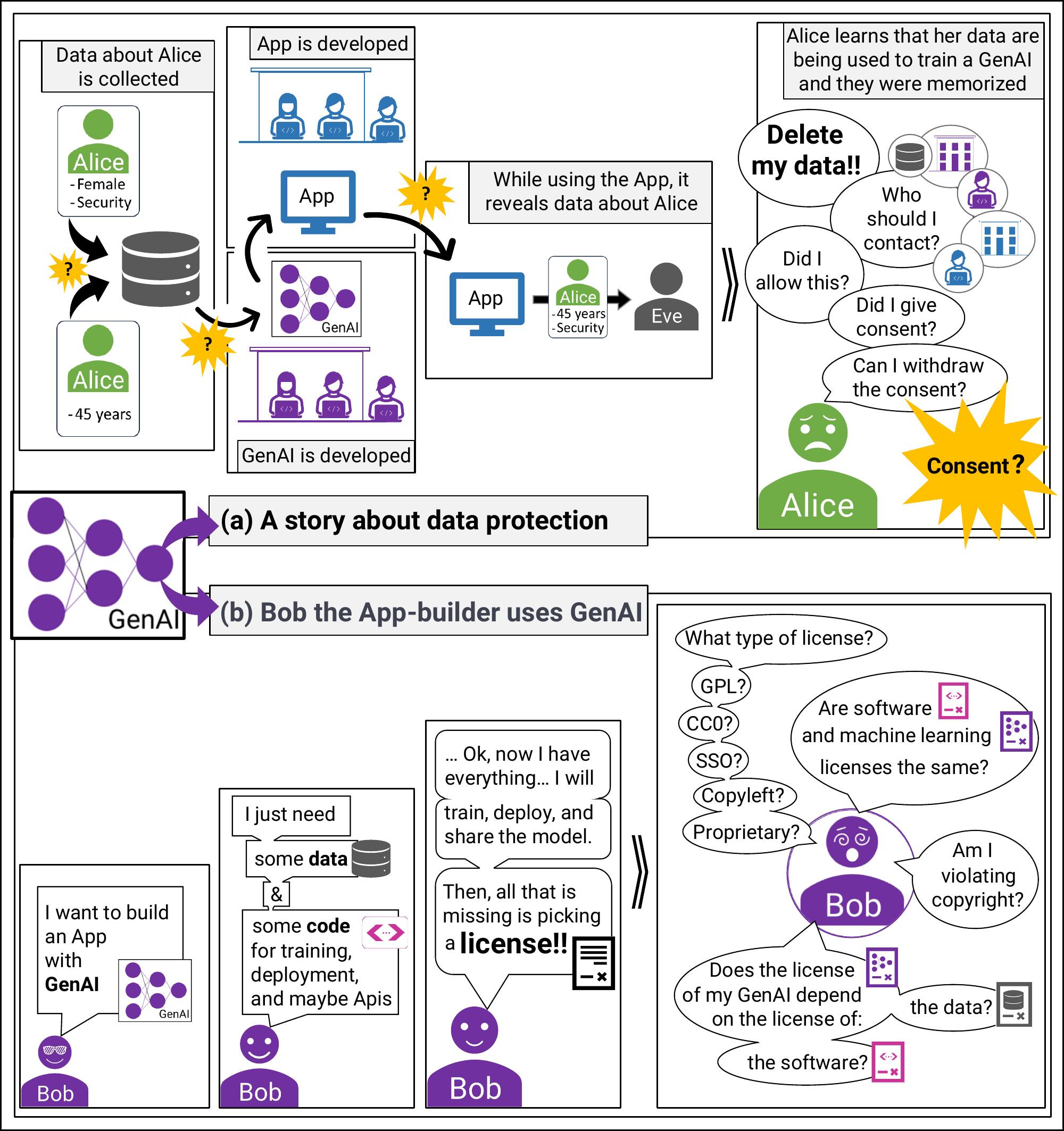}}
\caption{Alice and Bob in the world of GenAI.}\vspace*{-5pt}
\label{fig:joined-comic}
\end{figure*}

\section{Data Protection}

The General Data Protection Regulation (GDPR)\cite{GDPR2016a} was a milestone for data protection in the European Union and gave users the power to decide if and how their personal data may be used. Examples of protected personal data are names, locations, photos, beliefs, e-mail addresses, but also communication behavior (Art. 1 GDPR). The impact of this regulation cannot be overstated, and other data protection regulations often follow the GDPR model. For example,  the rules from the GDPR changed the global internet by requiring websites to obtain active consent for storing cookies, which led to the infamous cookie banners. Since other regulations like the CCPA are typically weaker than the GDPR, we focus our discussion on the GDPR. 

Figure~\ref{fig:joined-comic}(a) shows how personal data becomes part of GenAI models: Alice's data becomes part of a large data set that is used to train a GenAI model. This model can be used by the same or by different developers to create applications. We know hat these models sometimes memorize data and reveal them as part of their output.\cite{carlini2023quantifying} Obviously, this means that there is also a potential to memorize personal data and reveal them to other users like Bob. At this point, Alice may become aware of this use of her personal data and consider what her options are since she knows that personal data is protected under the GDPR. 

\subsection{Consent and legitimate interest}

A key question for what happens next is whether Alice gave consent for the processing of her personal data. Did Alice give consent for the collection of her data? Did her consent include the processing of her data for the training of a GenAI model? Does this consent also involve the publication of an application for a specific purpose? Does the consent allow the reproduction of her data? Even if Alice consented to have her data stored and processed (e.g., as part of the presence on a website), it is unclear if and how this consent propagates to the downstream use. The GDPR outlines several principles for the processing of personal data that clarify relevant restrictions (Art. 5 GDPR):

\begin{itemize}
    \item[{\ieeeguilsinglright}] The use of personal data must be \textit{transparent}, i.e., who processes the data with which type of data processing. This includes transparency about the scope of the processing and possible risks involved. 
    \item[{\ieeeguilsinglright}] The \textit{purpose limitation} means that all such processing must have a clearly specified purpose and that no further processing is allowed. This raises the question of whether it is valid to use any personal data for GenAI training unless consent for this use case was given. 
    \item[{\ieeeguilsinglright}] The \textit{data minimization} principle strengthens the purpose limitation and implies that data protection is the default, which means that usage of personal data is opt-in rather than opt-out.
\end{itemize}

From these general considerations, we can deduce the legal implications of the different possible scenarios we observe in Figure~\ref{fig:joined-comic}(a), depending on Alice's consent. If Alice consented specifically to the use of her data for the GenAI training and the subsequent use within the App, there are usually no legal problems. We say ``usually'' because Alice might also be a minor, considered younger than 16 years by the GDPR. In this case, consent must be obtained from the holder of parental responsibility (Art. 8 GDPR). 

In case there is no consent, the GDPR also allows several other options, including the \textit{legitimate interest} (Art. 6 GDPR).
The meaning of legitimate interest is not clearly defined by the GDPR and must be considered on a case-by-case basis (rec. 47 GDPR). To decide if something qualifies as a legitimate interest, you have to weigh Alice's desire to protect her data against the interests of the users of this data and whether Alice can reasonably expect that her data will be processed. In our case, the data users are the companies seeking to train and use GenAI models. According to 
an analysis by the European Parliament, training GenAI models is permissible as long as reasonable security measures, such as anonymization, are applied.\cite{gdprImpactAI} A broader reading of legitimate interest could also consider if it is technically possible to apply anonymization at all, e.g., when it comes to GenAI models for computer vision that require the use of Alice's image. Recently, published guidelines from the French Data Protection Authority (CNIL) also consider such special cases to be legitimate. \cite{cnilEnsuringLawfulness} Similarly, one might question whether anonymization is feasible when training large language models on large portions of the publicly available internet and whether this would also fall under legitimate interest; even without such steps to ensure the protection of personal data. 

The above considerations apply to the use of Alice's data for training GenAI models. However, GenAI models do not only learn an abstract representation of the training data, they generate content. This content is sometimes memorized from the training data.\cite{carlini2023quantifying} This alters the considerations of consent and legitimate interest. If a model is able to reveal personal data, consent must be obtained for this purpose. Crucially, this means: consent for all downstream applications that may reveal such personal data likely needs to be obtained separately. Specifically, if the fictional model in Figure~\ref{fig:joined-comic}(a) disclosed Alice's personal information to Eve and added insults, it would constitute a second violation of Alice's rights. For legitimate interests, this means that the use of data pseudonymization and anonymization become more important. 

If there is neither consent nor legitimate interest, the GDPR still allows the use of personal data if there is a \textit{public interest}. This is good news for researchers, as research is typically considered a public interest. Therefore, research on GenAI is usually on firm legal grounds.

The above discussion shows that relationship between personal data and GenAI is complex. Legitimate interest covers many uses of such data, but probably not if the personal data is memorized and can be revealed by the GenAI model. 

\begin{mdframed}
\textbf{Lesson 1. Many use cases are covered by the legitimate interest clause of the GDPR, as long as no memorization of the personal data happens. The legally safe method is either to obtain consent or to pseudonymize/anonymize data before training.}
\end{mdframed}

\subsection{Alice's rights}

This raises the question of what would happen if the processing was not covered by the GDPR, either through Alice's consent, the legitimate interest, or the public interest. Moreover, the GDPR awards Alice the right to withdraw her consent at any time. Since companies are not allowed to switch the legal basis of processing automatically, they cannot claim that further processing of Alice's data is allowed under the GDPR as a legitimate interest. In all cases when there is no such legal basis, the GDPR awards Alice various \textit{data subject rights} (Art. 16 GDPR). We examine two of these rights, i.e., the \textit{right to restriction of processing} (Art. 18 GDPR) and the \textit{right to erasure} (Art. 17 GDPR). 

The \textit{right to restriction of processing} means that Alice has the right to stop the use of her data. Restrictions like this would certainly affect the training of new GenAI models. The how much of an impact this has depends on the nature of the GenAI model. While this is rather an inconvenience for textual data, which can be filtered with regular expressions, it can be fairly difficult for other data like images and voice, which may be harder to detect automatically within training data. Ironically, preventing the GenAI from processing Alice's data may require training an AI model with her data to locate them first. However, this would likely fall under legitimate interest and be allowed under the GDPR.

However, restricting processing could also impact existing GenAI models. For instance, Italy's data protection authority restricted access to ChatGPT for a certain time, in part because they were uncertain about the legal basis of the data processing.\cite{italyChatGPT} Moreover, the current generation of GenAI models is known to hallucinate, i.e., makeup content. This can also affect personal data, which may then be inaccurately represented. The GDPR also requires the processing of personal data to be accurate (Art. 5 GDPR). This is a practical issue beyond the GDPR is, e.g., highlighted by the case of a major in Australia who fights generated misinformation.\cite{washingtonpostChatGPTInvented}

The question then arises of how software developers could deal with processing restrictions for already existing models. In principle, there are two solutions. The first is to post-process the output of the GenAI to filter personal data from Alice, which would be similar to the filters required for training data. This \textit{privacy box} would be similar to \textit{shielding}\cite{Alshiekh_Bloem_Ehlers_Könighofer_Niekum_Topcu_2018}, which is a solution to post-process AI output to make it safe. The second and cleaner solution is more difficult. It would involve modifying the GenAI model so that it can no longer generate Alice's personal data. However, current technology cannot reliably make such modifications as \textit{machine unlearning} remains an unresolved issue.\cite{unlearningchallengeNeurIPS2023}

Another question is who would be responsible for enforcing the processing restriction. In our scenario in Figure~\ref{fig:joined-comic}(b), the PURPLE company develops the GenAI model, and the BLUE company develops the app. In real life, this could mean that the app developed by BLUE consumes an API provided by PURPLE. This situation is further complicated when we consider that the legitimate interests for using Alice's data may differ between PURPLE and BLUE and that Alice might not restrict all processing of her data, but only specifically for BLUE's app because she dislikes their business case. Consequently, determining which companies need to address this issue requires a case-by-case analysis. 

The \textit{right to erasure} goes beyond simply ceasing the processing of Alice's data and also requires the deletion of all copies of her data. For the training data, this means that instead of filtering Alice's data, all instances of her data must be directly deleted from the dataset. Consequently, the technical challenges here are very similar to those of processing restrictions, and the key difference is that dynamic filtering at processing time is no longer sufficient.

To be able to enforce the right to erasure for the GenAI model, Alice must demonstrate that her data was memorized and could be revealed. Thus, if the GenAI model only used Alice's data during training, but there is no prompt yet that can demonstrate that Alice's data is memorized, the right to erasure does not apply to the GenAI model. If Alice can demonstrate how her data is revealed by the model, the provider of the GenAI model has to delete this. 

However, as previously discussed, machine unlearning is not yet feasible. Therefore, the only way to achieve this is by retraining the model without Alice's data, which may require a significant effort. Fortunately for GenAI developers, the right to erasure is not absolute and it may also be an option to put data \textit{beyond use} with a privacy box, as described above. An example of this is backups: while personal data needs to be erased from current working copies upon request, it is not always necessary to do so from backups, as long as they are not used and the effort for deletion would be significant. Additionally, there needs to be a schedule for when the backup will be replaced.\cite{icoRightErasure} While the scenario for GenAI is different, one could argue that unlearning is technically infeasible to unlearn without damaging other data and that retraining is infeasible. In such cases, a \emph{beyond-use} option may be considered as an alternative, provided that the data is no longer accessible and will be deleted the next time this is reasonably possible, i.e., when a re-training is scheduled. 

In summary, compliance with data subject rights is fairly straightforward for the training data, even though there is an interesting corner case in which an AI model needs to be used to detect personal data. As long as machine unlearning is infeasible, it should suffice to place models in a privacy box that prevents the disclosure of personal data. 

\begin{mdframed}
\textbf{Lesson 2. Filtering mechanisms are a viable safeguard to ensure the rights of data subjects when they request that their data no longer be processed or stored.}
\end{mdframed}

\subsection{Understanding the risks}

Overall, the above analysis shows that the GDPR does not really restrict the rise of GenAI technologies as long as providers of such technologies are prepared to filter out personal data upon request. However, this has not yet been tested in the European courts, which may mean that our analysis -- especially regarding legitimate interest -- may be incorrect. Still, this is not unknown from other GDPR-related aspects like the cookie banners and would likely only require that future changes follow the guidance established by such rulings. 

In case subject rights are violated because data are used even when consent was withdrawn or without a legitimate interest (e.g., minors are involved), the GDPR allows compensation for any damage suffered (Art. 82 GDPR). Thus, illegal processing without damages only needs to be stopped, but if Alice can show concrete material of psychological damages, there would be legal exposure for payment of compensation. 

\section{Licenses for GenAI}

GenAI models do not magically appear from thin air. Instead, they are trained with a software that consumes training data to learn to generate  content. Whenever we reuse something, we need a legal foundation for this, which is where the licenses become relevant. As software developers are aware, licensing questions can be tricky. Many companies are wary when re-using publicly available software, e.g., to avoid accidental use of code with a copyleft license (e.g., GPL). 

Figure~\ref{fig:joined-comic}(b) shows a common scenario where Bob wants to create a GenAI model. Typically, the data is not created by Bob but rather collected from somewhere. Similarly, training code is typically at least partially reused from prior work, e.g., research papers or from publicly available models. Both data and code were published by someone who automatically got the copyright for this content. Through licenses, they can allow others to reuse this content. Bob needs to know about these licenses, as well as understand how licenses for GenAI models differ from software licenses. 

For our following considerations, we consider four categories of licenses: 1) The \textit{public domain} that allows unrestricted use, e.g., when content is no longer covered by copyright due to age or specifically placed there with a license like CC0. 2) \textit{Permissive} licenses that allow the reuse of data and code without major restrictions, such as the MIT License, the Apache License, or the CC-BY license. 3) \textit{Copyleft} licenses allow the reuse of the licensed work but enforce that the same type of license is used for derivative work. 4) \textit{Proprietary} licenses typically do not grant reuse unless there is compensation. 

\subsection{Model licenses}

Before considering which license Bob can use for his GenAI, he needs to know how model licenses differ from software licenses. Generally, both are similar and regulate how the software and the GenAI model may be used. While it is possible to use a software license for GenAI models, model licenses may also additionally regulate machine learning-specific aspects. Examples of such licenses are the OpenRAIL licenses. OpenRAIL is, in general, permissive and modeled after the Apache License but requires that downstream use is responsible and, thereby, excludes certain use cases like the generation of misinformation. Notably, the OpenRAIL licenses allow these limitations to be placed on the training data, the application, the model, and/or the source code, i.e., the license allows a fine-grained selection of which aspects of the GenAI model need to be re-used responsibly. From this, we have the following message for Bob and other software developers:

\begin{mdframed}
\textbf{Lesson 3: Unless you want to restrict the use of your GenAI model, you can use normal software licenses. To restrict downstream use of specific aspects of your GenAI model, machine learning licenses can be used, but they are still emerging.}
\end{mdframed}

\subsection{Weak and Strong Copyleft}

Another very important consideration is whether strong or weak copyleft licenses are used. Weak copyleft licenses (e.g., LGPL, MPL, EPL) only require the sharing of modified aspects. In our case, this would mean modified data, modified models, or modified training code. 

With a strong copyleft license (e.g., GPL, CC-BY-SA), the restriction is stronger and typically requires sharing the complete derivative work. Arguably, this would mean that if the data, model, or training code used to build a GenAI is built under a copyleft license, then the resulting product, including the model itself, the training code, and any new data, must be licensed under a compatible copyleft license. Such a publication can possibly be problematic, e.g., if the license for the data does not permit this. To the best of our knowledge, no such license currently allows for such distinctions, meaning that copyleft licenses may be problematic for GenAI, if they are strictly enforced, including for training data from downstream use. 

\begin{mdframed}
\textbf{Lesson 4. Be wary of strong copyleft licenses when working with GenAI. You may need to share more than you want to or are legally allowed to.}
\end{mdframed}

\subsection{Ignoring licenses}

Bob may also do what many current actors in the GenAI space are doing: ignore the licenses for the data and the models. For the data, this is typically argued based on fair use or similar exceptions from copyright. Fair use allows the use of copyrighted material under certain conditions, including as inspiration for creative processes. The argument is that GenAI is a creative process that is only inspired by the training data but does not copy it. Currently, ongoing litigation in the USA supports this legal argument but also leaves the door open for copyright claims, in case it can be shown that a GenAI model memorized the licensed data. A recent experiment with Midjourney and Dall-e 3, as well as a lawsuit by the New York Times that provides such examples, indicate that this battle over data licenses is not yet over.\cite{ieeeGenerativeVisual} In the European Union, the Digital Single Markets Act (DSMA) provides exceptions that allow ignoring copyright for data mining for research already unless explicitly forbidden. Some countries, like Germany, have also granted such an exception for commercial purposes. However, in the same manner that the litigation in the US left the door open for copyright claims, the European exceptions only cover the training, while generated content that would be a copy of training data may still infringe copyright. 

For the models, it is unclear if we are allowed to ignore licenses. The legal question is completely different than for the data, as the argument is not fair use but rather whether models are copyrightable. As a legal principle, copyright can only be awarded if a human creative aspect is involved, which generally excludes results from the computation of an algorithm. This also applies to GenAI. However, in reality, humans are often involved and do make creative decisions that affect the model weights, e.g., selecting data, curating data, tuning algorithms, and babysitting training, all of which are reasons that models may indeed be copyrightable. How this will play out is currently completely uncertain and, e.g., considered actively by the US copyright office.\cite{copyright}

\begin{mdframed}
\textbf{Lesson 5. You probably can ignore copyright and licenses when training GenAI models for research, as long as it does not memorize training data to produce material that infringes copyright when generating. For commercial purposes, you need to check the local rules, and licenses may be required. For models, we suggest it is better to be safe than sorry and follow the licenses.}
\end{mdframed}

\section{CONCLUSION}

Overall, the situation surrounding software development with GenAI is fairly simple: if the GenAI model does not memorize personal data or licensed data, there are no problems. If such data is memorized, it must be prevented from being revealed unless consent or appropriate license agreements are in place. The implications of copy-left for GenAI models should be considered when placing models under such licenses, as this may require disclosure of data when retraining such models, indicating a need for more specialized licenses. However, since all of this is relatively new, there are only few cases where these legal interpretations have been tested in the courts, which means there is still some residual legal risk, even if developers follow our lessons.

\def\refname{REFERENCES}


\bibliographystyle{IEEEtran}
\bibliography{bibliography}

\begin{thebibliography}{10}
\providecommand{\url}[1]{#1}
\csname url@samestyle\endcsname
\providecommand{\newblock}{\relax}
\providecommand{\bibinfo}[2]{#2}
\providecommand{\BIBentrySTDinterwordspacing}{\spaceskip=0pt\relax}
\providecommand{\BIBentryALTinterwordstretchfactor}{4}
\providecommand{\BIBentryALTinterwordspacing}{\spaceskip=\fontdimen2\font plus
\BIBentryALTinterwordstretchfactor\fontdimen3\font minus \fontdimen4\font\relax}
\providecommand{\BIBforeignlanguage}[2]{{%
\expandafter\ifx\csname l@#1\endcsname\relax
\typeout{** WARNING: IEEEtran.bst: No hyphenation pattern has been}%
\typeout{** loaded for the language `#1'. Using the pattern for}%
\typeout{** the default language instead.}%
\else
\language=\csname l@#1\endcsname
\fi
#2}}
\providecommand{\BIBdecl}{\relax}
\BIBdecl

\bibitem{GDPR2016a}
\BIBentryALTinterwordspacing
{European Parliament} and {Council of the European Union}. Regulation ({EU}) 2016/679 on the protection of natural persons with regard to the processing of personal data and on the free movement of such data, and repealing {Directive} 95/46/{EC} ({General} {Data} {Protection}). [Online]. Available: \url{https://data.europa.eu/eli/reg/2016/679/oj}
\BIBentrySTDinterwordspacing

\bibitem{carlini2023quantifying}
\BIBentryALTinterwordspacing
N.~Carlini, D.~Ippolito, M.~Jagielski, K.~Lee, F.~Tramer, and C.~Zhang, ``Quantifying memorization across neural language models,'' in \emph{11th International Conference on Learning Representations}, 2023. [Online]. Available: \url{https://openreview.net/forum?id=TatRHT_1cK}
\BIBentrySTDinterwordspacing

\bibitem{gdprImpactAI}
\BIBentryALTinterwordspacing
{European Parliamentary Research Service}. {The impact of the General Data Protection Regulation on artificial intelligence}. [Online]. Available: \url{https://www.europarl.europa.eu/RegData/etudes/STUD/2020/641530/EPRS_STU(2020)641530_EN.pdf}
\BIBentrySTDinterwordspacing

\bibitem{cnilEnsuringLawfulness}
\BIBentryALTinterwordspacing
{Commission Nationale de l'Informatique et des Libertés}. {E}nsuring the lawfulness of the data processing. [Online]. Available: \url{https://cnil.fr/en/ensuring-lawfulness-data-processing}
\BIBentrySTDinterwordspacing

\bibitem{italyChatGPT}
\BIBentryALTinterwordspacing
{Italy curbs ChatGPT, starts probe over privacy concerns}. [Online]. Available: \url{https://www.reuters.com/technology/italy-data-protection-agency-opens-chatgpt-probe-privacy-concerns-2023-03-31}
\BIBentrySTDinterwordspacing

\bibitem{washingtonpostChatGPTInvented}
\BIBentryALTinterwordspacing
{C}hat{G}{P}{T} invented a sexual harassment scandal and named a real law prof as the accused. [Online]. Available: \url{https://www.washingtonpost.com/technology/2023/04/05/chatgpt-lies/}
\BIBentrySTDinterwordspacing

\bibitem{Alshiekh_Bloem_Ehlers_Könighofer_Niekum_Topcu_2018}
\BIBentryALTinterwordspacing
M.~Alshiekh, R.~Bloem, R.~Ehlers, B.~Könighofer, S.~Niekum, and U.~Topcu, ``Safe reinforcement learning via shielding,'' in \emph{AAAI Conference on Artificial Intelligence}, 2018. [Online]. Available: \url{https://doi.org/10.1609/aaai.v32i1.11797}
\BIBentrySTDinterwordspacing

\bibitem{unlearningchallengeNeurIPS2023}
\BIBentryALTinterwordspacing
{N}eur{I}{P}{S} 2023 {M}achine {U}nlearning {C}hallenge --- unlearning-challenge.github.io. [Online]. Available: \url{https://unlearning-challenge.github.io/}
\BIBentrySTDinterwordspacing

\bibitem{icoRightErasure}
\BIBentryALTinterwordspacing
I.~C. Office. {R}ight to erasure. [Online]. Available: \url{https://ico.org.uk/for-organisations/uk-gdpr-guidance-and-resources/individual-rights/individual-rights/right-to-erasure/#ib5}
\BIBentrySTDinterwordspacing

\bibitem{ieeeGenerativeVisual}
\BIBentryALTinterwordspacing
G.~Marcus and R.~Southen. {G}enerative {A}{I} {H}as a {V}isual {P}lagiarism {P}roblem. [Online]. Available: \url{https://spectrum.ieee.org/midjourney-copyright}
\BIBentrySTDinterwordspacing

\bibitem{copyright}
\BIBentryALTinterwordspacing
{United States Copyright Office}. Artificial intelligence and copyright. [Online]. Available: \url{https://www.copyright.gov/ai/docs/Federal-Register-Document-Artificial-Intelligence-and-Copyright-NOI.pdf}
\BIBentrySTDinterwordspacing

\end{thebibliography}

\begin{IEEEbiography}{Steffen Herbold}{\,} is the Chair for AI Engineering at University of Passau at Passau, Germany. His research interests include the quality and operation of machine learning. Herbold received his PhD in computer science from the University of Goettingen, Germany. He is a lifetime member of the ACM. Contact him at steffen.herbold@uni-passau.de.
\end{IEEEbiography}

\begin{IEEEbiography}{Brian Valerius}{\,} is the Chair for AI in Criminal Justice at the University of Passau at Passau, Germany. His research interests are on the impact of AI on the legal practice. Valerius received his J.D. from the University of Wuerzburg, Germany. Contact him at brian.valerius@uni-passau.de.
\end{IEEEbiography}

\begin{IEEEbiography}{Anamaria Mojica-Hanke}{\,} is a doctoral student at the University of Passau at Passau, Germany. Her research interests are machine learning aplication and practices for software engineering. Mojica-Hanke received her Master of Information Engineering from Universidad de los Andes, Colombia. Contact her at anamaria.mojica-hanke@uni-passau.de.
\end{IEEEbiography}

\begin{IEEEbiography}{Isabella Lex}{\,} is a doctoral student at the University of Passau at Passau, Germany. Her research interests are on the impact of AI on legal practice. Lex studied law at the University of Passau, Germany. Contact her at isabella.lex@uni-passau.de.
\end{IEEEbiography}

\begin{IEEEbiography}{Joel Mittel}{\,} is a doctoral student at the University of Passau at Passau, Germany. His research interests are on the impact of AI on legal practice. Mittel studied law at the University of Passau, Germany. Contact him at joel.mittel@uni-passau.de.
\end{IEEEbiography}



\end{document}